\documentclass[aps,pra, twocolumn, amsmath, amssymb, showpacs, superscriptaddress]{revtex4-2}

\usepackage{graphicx}
\usepackage{bm}

\usepackage{xcolor}

\begin{document}

\title{Optical clock based on two-photon spectroscopy of the nuclear transition in ion $^{229}$Th in a monochromatic field}

\author{V.\,I.\,Yudin}
\email{viyudin@mail.ru}
\affiliation{Novosibirsk State University, Novosibirsk, 630090 Russia}
\affiliation{Institute of Laser Physics, Siberian Branch, Russian Academy of Sciences, Novosibirsk, 630090 Russia}
\affiliation{Novosibirsk State Technical University, Novosibirsk, 630073 Russia}
\author{A.\,V.\,Taichenachev}
\affiliation{Novosibirsk State University, Novosibirsk, 630090 Russia}
\affiliation{Institute of Laser Physics, Siberian Branch, Russian Academy of Sciences, Novosibirsk, 630090 Russia}
\author{O.\,N.\,Prudnikov}
\affiliation{Novosibirsk State University, Novosibirsk, 630090 Russia}
\affiliation{Institute of Laser Physics, Siberian Branch, Russian Academy of Sciences, Novosibirsk, 630090 Russia}
\author{M.\,Yu.\,Basalaev}
\affiliation{Novosibirsk State University, Novosibirsk, 630090 Russia}
\affiliation{Institute of Laser Physics, Siberian Branch, Russian Academy of Sciences, Novosibirsk, 630090 Russia}
\affiliation{Novosibirsk State Technical University, Novosibirsk, 630073 Russia}
\author{A.\,N.\,Goncharov}
\affiliation{Novosibirsk State University, Novosibirsk, 630090 Russia}
\affiliation{Institute of Laser Physics, Siberian Branch, Russian Academy of Sciences, Novosibirsk, 630090 Russia}
\author{S.\,V.\,Chepurov}
\affiliation{Institute of Laser Physics, Siberian Branch, Russian Academy of Sciences, Novosibirsk, 630090 Russia}
\author{V.\,G.\,Pal'chikov}
\affiliation{All-Russian Research Institute of Physical and Radio Engineering Measurements, Mendeleevo, Moscow region, 141570 Russian}
\affiliation{National Research Nuclear University MEPhI (Moscow Engineering Physics Institute), Moscow, 115409 Russia}

\begin{abstract}
For the isotope $^{229}$Th we investigate the possibility of two-photon laser spectroscopy of the nuclear clock transition (148.38~nm) using intense monochromatic laser field at twice the wavelength (296.76~nm). Our estimates show that due to the electron bridge process in the doubly ionized ion $^{229}$Th$^{2+}$ the sufficient intensity of a continuous laser field is about 10-100\,kW/cm$^2$, which is within the reach of modern laser systems. This unique possibility is an result of the presence in the electronic spectrum of the ion $^{229}$Th$^{2+}$ of an exceptionally close intermediate (for the two-photon transition) energy level, forming a strong dipole ($E1$) transition with the ground state at the wavelength of 297.86~nm, which differs from the probe field wavelength (296.76~nm) by only 1.1~nm. The obtained results can be used for the practical creation of ultra-precise nuclear optical clocks based on thorium-229 ions without using vacuum ultraviolet.\\
Moreover, we develop an alternative approach to the description of the electron bridge phenomenon in an isolated ion (atom) using the hyperfine interaction operator, that is important for the general quantum theory of an atom. In particular, this approach shows that the contribution to the electron bridge from the nuclear quadrupole moment can be comparable to the contribution from the nuclear magnetic moment.
\end{abstract}

\maketitle

It is common knowledge \cite{Riehle_2005} that time and frequency are the most accurately measured physical quantities. Modern optical frequency standards (optical atomic clocks) based on ensembles of ultracold atoms in optical lattices and single ions in rf traps have achieved the fractional instability and uncertainty at the level of 10$^{-18}$ (and even a little better) \cite{Hinkley_2013,Huntemann_2016,Brewer_2019,Aeppli_2024}. Promising directions related to multi-ion systems \cite{Hausser_2024} and high-charged ions \cite{Kozlov_2018,Lyu_2025} are being successfully developed.

However, due to the recent experimental achievements and theoretical analysis, it seems very likely that the next decisive step to improve the metrological characteristics of quantum frequency standards will be made on the basis of the transition in the nucleus of the $^{229}$Th isotope between the ground state and the low-lying isomeric state with energy 8.4~eV. As is known (see \cite{Tkalya_2003} and references cited therein), the first indications on the existence of a low-lying isomeric state in  $^{229}$Th were obtained in the 70s of the 20th century. The first proposal to excite the isomeric state in $^{229}$Th by optical photons through the electron bridge process belongs to E.\,V.\,Tkalya \cite{Tkalya_1992}. The pioneering proposal to create an ultra-precision optical clock on the basis of this transition was made by E.~Peik and Chr.~Tamm in \cite{Peik_2003}. The basic idea of this paper was that the clock frequency in the nucleus is significantly less sensitive to changes in external environment compared to the frequency of any transition in the electron shell of an atom. Already then, the possibility of direct excitation of the nuclear transition by radiation from an ultra-stable laser or a femtosecond laser comb and detection of the transition by fluorescence or by a change in the magnitude of hyperfine splitting was noted. Different variants of the precision spectroscopy were discussed: in rf ion traps, using laser cooling, or in transparent crystals. In \cite{Campbell_2012}, the fractional uncertainty of the frequency standard based on trapped and laser-cooled $^{229}$Th$^{3+}$ ions was estimated at the level of 10$^{-19}$. The possibilities of achieving similarly high metrological characteristics in solid-state nuclear clocks were investigated in \cite{Kazakov_2012}.

In the period 1990-2020, the accuracy of determining the transition energy between the ground and isomeric states in the $^{229}$Th nucleus in various experiments \cite{Helmer_2005,Filho_2005,Beck_2007,Beck_2009,Seiferle_2019,Yamaguchi_2019,Sikorsky_2020} gradually increased until it reached in 2023 the level of 8.338(24)~eV \cite{Kraemer_2023}, sufficient for direct excitation of the transition by high-power pulsed lasers. The first experimental observation of laser excitation of the isomeric state in the thorium-229 nucleus was published in \cite{Okhapkin_2024}. This was soon followed by measurements \cite{Elwell_2024,Ye_2024,Zhang_2024}, in which the results of \cite{Okhapkin_2024} were confirmed and improved using other solid-state matrices and films, containing thorium-229, and alternative laser sources at wavelength 148.4~nm. Thus, the frequency of the nuclear transition in $^{229}$Th was determined with kilohertz precision, which opens the way to the practical creation of nuclear optical clocks.

However, one of the main scientific-technical problems that still need to be solved is the creation of precision laser systems for interrogation of the clock nuclear transition in vacuum ultraviolet (VUV) 148.4~nm. At least two alternative approaches are being developed in this direction: (A) femtosecond frequency comb in the VUV based on a powerful and precise femtosecond synthesizer in the infrared range and generation of higher harmonics in a suitable medium; (B) a precision single-frequency continuous laser in the VUV based on semiconductor emitters using high-frequency resonators, amplifiers, and nonlinear frequency transformations in crystals. Both approaches encounter significant difficulties, apparently of a technical nature. For example, the femtosecond comb implemented in \cite{Ye_2024} is not powerful enough or precise enough to interrogate the clock transition. At the same time, for the single-frequency approach, nonlinear crystals with proven required efficiency of second harmonic generation at the last step (from 296.8~nm to 148.4~nm) have not yet been realized \cite{Ohapkin}.

In this paper, we propose an alternative approach based on the possibility of two-photon laser spectroscopy of the nuclear clock transition using intense monochromatic laser radiation at 296.76~nm. Our estimates show that due to the electron bridge mechanism in the doubly ionized ion $^{229}$Th$^{2+}$ a sufficient intensity of continuous laser field is of the order of 10-100~kW/cm$^2$, which lies within the reach of modern laser systems. For example, a laser beam with a power of 1~W, focused in a diameter of 100~$\mu$m, gives an intensity of about 10~kW/cm$^2$. It should be noted that two-photon excitation by two-frequency and polychromatic laser radiation of the isomeric state in the thorium-229 nucleus has been repeatedly discussed by various authors \cite{Porsev_2010,Muller_2019,Borisyuk_2019,Cai_2021,Karpeshin_2024} in the context of searching for the clock transition and accurately determining its frequency. However, the use of the two-photon spectroscopy in a monochromatic field to interrogate the nuclear clock transition in $^{229}$Th and create an ultra-precise nuclear clock is proposed, to our knowledge, for the first time.

We will analyse the possibility of two-photon spectroscopy of the nuclear clock transition in $^{229}$Th ions through electronic states. Such a principal possibility exists due to the so-called electron bridge, which can mix the energy states of an atom (ion) with different nuclear states. In the case of the isotope $^{229}$Th, the electron bridge mixes the energy levels of the ion with nuclear spin in the ground state 5/2 and the first excited state of the isomer with nuclear spin 3/2. In \cite{Krutov_1968,Hinneburg_1979,Hinneburg_1981,Tkalya_1991,Tkalya_1992}, the electron bridge was described in the framework of quantum electrodynamics using the Feynman diagram technique. However, as we show below, the electron bridge mechanism can also be described in terms of standard quantum atomic physics using the hyperfine interaction operator, which consists of two contributions
\begin{equation}\label{hfs_gen}
\hat{V}^{(\rm hf)}=\hat{V}^{(\rm hf)}_{\bf B}+\hat{V}^{(\rm hf)}_{\rm Q},
\end{equation}
where the first contribution is due to the nuclear magnetic moment and the second contribution is due to the nuclear quadrupole moment.

Let us consider the nucleus as a source of the vector potential $\hat{{\bf A}}_{\rm n}({\bf r})$, which forms the corresponding magnetic field $\hat{{\bf B}}_{\rm n}({\bf r})$=$[\nabla$$\times$$\hat{{\bf A}}_{\rm n}({\bf r})]$, where the symbol $[{\bf a}$$\times$${\bf b}]$ means the vector product of vectors ${\bf a}$ and ${\bf b}$, and $\nabla$=${\bf e}_{x}\partial/\partial x$+${\bf e}_{y}\partial/\partial y$+${\bf e}_{z}\partial/\partial z$ is the standard gradient operator. Then, according to the known formulas of quantum atomic physics, we have the following general expression for the magnetic contribution
\begin{equation}\label{hfs_mu_gen}
\hat{V}^{(\rm hf)}_{\bf B}=2\mu_{\rm B}\sum^{}_{j}\big\{\hbar^{-1}\big(\hat{{\bf A}}_{\rm n}({\bf r}_{j})\cdot\hat{{\bf p}}_{j}\big)+\big(\hat{{\bf B}}_{\rm n}({\bf r}_{j})\cdot\hat{{\bf s}}_{j}\big)\big\},
\end{equation}
where ${\bf r}_{j}$, $\hat{{\bf p}}_{j}$, and $\hat{{\bf s}}_{j}$ are, respectively, the radius vector, the momentum operator, and the dimensionless (in units of $\hbar$) spin operator of the $j$-th electron, $\mu_{\rm B}=\hbar |e|/(2m_{e}c)$ is the Bohr magneton ($e$ and $m_{e}$ are the charge and mass of the electron, $c$ is the speed of light).

Usually, with a good accuracy, the nucleus can be considered as a point magnetic dipole characterized by the magnetic moment operator $\hat{\bm{\mu}}$, which forms the following vector-potential
\begin{equation}\label{A}
 \hat{{\bf A}}_{\rm n}({\bf r})=-[\hat{\bm{\mu}}\times\nabla] \frac{1}{|{\bf r}|}=\frac{[\hat{\bm{\mu}}\times {\bf r}]}{|{\bf r}|^3}.
\end{equation}
Using the known formulas for the change of the coupling scheme in the tensor product of three tensors \cite{Varshalovich_1988}, the magnetic field generated by the nucleus can be represented as
\begin{eqnarray}\label{B}
\hat{{\bf B}}_{\rm n}({\bf r})&=&[\nabla\times\hat{{\bf A}}_{\rm n}({\bf r})]=-[\nabla\times[\hat{\bm{\mu}}\times\nabla]]\frac{1}{|{\bf r}|}=\nonumber\\
&&\bigg[-\frac{2}{3}\hat{\bm{\mu}}\Delta -\frac{\sqrt{5}}{\sqrt{3}}\{\hat{\bm{\mu}}\otimes\{\nabla\otimes\nabla\}^{}_2\}^{}_1\bigg]\frac{1}{|{\bf r}|}=\nonumber\\
&&\frac{8\pi}{3}\delta({\bf r})\hat{\bm{\mu}}-\frac{\sqrt{15}\,\{\hat{\bm{\mu}}\otimes\{{\bf r}\otimes{\bf r}\}^{}_2\}^{}_1}{|{\bf r}|^5}=\nonumber\\
&&\frac{8\pi}{3}\delta({\bf r})\hat{\bm{\mu}}+\frac{3{\bf r}({\bf r}\cdot\hat{\bm{\mu}})-|{\bf r}|^2\hat{\bm{\mu}}}{|{\bf r}|^5},
\end{eqnarray}
where $\Delta$=$(\nabla$$\cdot$$\nabla)$ is the Laplacian, the symbol $\{{\cal A}^{}_{\kappa_1}$$\otimes$\,${\cal B}^{}_{\kappa_2}\}^{}_{\kappa}$ denotes the tensor product of the rank $\kappa$ of the two tensors ${\cal A}^{}_{\kappa_1}$ and ${\cal B}^{}_{\kappa_2}$ of ranks $\kappa_1$ and $\kappa_2$ (see \cite{Varshalovich_1988}), and	$\delta({\bf r})$ is the Dirac delta function [in (4) we used $\Delta$(1/$|{\bf r}|$)=$-4\pi$$\delta({\bf r})$].

Substituting the expressions (\ref{A})-(\ref{B}) in (\ref{hfs_mu_gen}) and performing simple mathematical transformations, the operator $\hat{V}^{(\rm hf)}_{\bf B}$ can be represented as a scalar product
\begin{equation}\label{hfs_mu}
\hat{V}^{(\rm hf)}_{\bf B}=(\hat{\bm{\mu}}\cdot\hat{{\bf K}}),
\end{equation}
in which the vector operator $\hat{{\bf K}}$, depending only on variables of electrons, has the form
\begin{equation}\label{K}
\hat{{\bf K}}=2\mu_{\rm B}\sum^{}_{j}\bigg[\frac{8\pi}{3}\delta({\bf r}_j)\hat{{\bf s}}_{j}+\\
\frac{\hat{{\bf l}}_{j}+3{\bf n}_j({\bf n}_j\cdot\hat{{\bf s}}_{j})-\hat{{\bf s}}_{j}}{|{\bf r}_{j}|^3}\bigg],
\end{equation}
where ${\bf n}_{j}$=${\bf r}_{j}/|{\bf r}_{j}|$ and $\hat{{\bf l}}_{j}$=$[{\bf r}_j\times\hat{{\bf p}}_{j}]/\hbar$ are, respectively, the unit vector of the radius-vector direction and the dimensionless (in units of $\hbar$) operator of the orbital angular momentum of the $j$-th electron. Note that the first term in (\ref{K}), containing $\delta({\bf r}_j)$, gives the contribution to the hyperfine interaction, which is due only to the presence of $s$-electrons in the unenclosed electron shells that form the energy structure of the ion (atom), since the wave functions of electrons with orbital momentum $l\geq 1$ is zero in the center $|{\bf r}_{j}|=0$. While the second term in (\ref{K}) contains the contributions to the hyperfine interaction from all electrons.

Let us show that the operator $\hat{V}^{(\rm hf)}_{\bf B}$, represented in the form (\ref{hfs_mu}), also allows to describe the connection between the states of an ion (atom) with different nucleus spin $I$, i.e. the electron bridge. For this purpose, let us consider the energy structure of the ion (atom), each state of which is described by the wave function $|F,m_{F},(n,J),I\rangle$, where $F$ is the total angular momentum ($|J-I|$\,$\leq$\,$F$\,$\leq$\,$J+I$), $m_{F}$ is the projection of the angular momentum on the quantization axis ($-F$\,$\leq$\,$m_{F}$\,$\leq$\,$F$), $J$ is the electronic angular momentum, and $I$ is the nucleus spin, which in general can have different values (as, for example, in the isotope $^{229}$Th). In addition, we have introduced the characteristic $n$, which identifies the electronic structure of the level under consideration with a given angular momentum $J$. In this basis, the magnetic contribution (\ref{hfs_mu}) to the hyperfine interaction operator is determined by matrix elements, which, in accordance with the known formulas of the quantum theory of angular momentum (see \cite{Varshalovich_1988}), have the form
\begin{eqnarray}\label{matr_el}
&& \langle F',m_{F'},(n',J'),I'| \hat{V}^{(\rm hf)}_{\bf B}|F,m^{}_{F},(n,J),I\rangle =\\
&& \delta^{}_{F'F}\delta^{}_{m_{F'}m_{F}}(-1)^{F+I+J'}
\left\{\begin{array}{ccc}
I' & I & 1 \\
J& J' & F \\
\end{array}
\right\}
\mu^{}_{I',I}K^{}_{n'J',nJ}\,, \nonumber
\end{eqnarray}
where $\mu^{}_{I',I}$$\equiv$$\langle I'||\hat{\bm{\mu}}||I\rangle$ is the reduced matrix element of the nuclear magnetic moment between the nucleus states $|I'\rangle$ and $|I\rangle$ with spins $I'$ and $I$, while $K^{}_{n'J',nJ}$$\equiv$$\langle n',J'||\hat{{\bf K}}||n,J\rangle$ is the reduced matrix element of the operator (\ref{K}) between the electronic states $|n',J'\rangle$ and $|n,J\rangle$. The electronic states and nuclear states must have in pairs the same parity, and their angular momenta can differ no more than one: $|J-J'|\leq 1$, $|I-I'|\leq 1$.

In the case of diagonal matrix elements in (\ref{matr_el}), the value
\begin{equation}\label{hfs_shift}
\Delta^{(\mu )}_{FnJI}=
(-1)^{F+I+J}
\left\{\begin{array}{ccc}
I & I & 1 \\
J& J & F \\
\end{array}
\right\}
\mu^{}_{I,I}K^{}_{nJ,nJ}\,,
\end{equation}
describes the hyperfine shift of the corresponding hyperfine level due to magnetic interactions. In this case, the magnetic moment of a nucleus with a fixed value of spin $I$ is usually defined as
\begin{equation}\label{mu_nuc}
\hat{\bm{\mu}}=g^{}_I\mu^{}_{\rm N}\hat{{\bf I}},
\end{equation}
where $\hat{{\bf I}}$ is the spin operator of the nucleus, $\mu^{}_{\rm N}$=$\hbar |e|/(2m_{p}c)$ is the nuclear magneton ($m_{p}$ is the proton mass), and $g^{}_I$ is the nuclear $g$-factor, the value of which is measured experimentally. Then we obtain for the diagonal reduced matrix elements
\begin{equation}\label{diag_el_mu}
\mu^{}_{I,I}=g^{}_I\mu^{}_{\rm N}\sqrt{I(I+1)(2I+1)}\,,
\end{equation}
following the standard definition (e.g. see \cite{Varshalovich_1988}).

In the case of different nuclear states (e.g., $I$\,$\neq$\,$I'$), the matrix elements in (\ref{matr_el}) describe the entanglement of states with different nuclear spins, which is the essence of the so-called electronic bridge. In particular, if the magnitude of the magnetic contribution to the hyperfine shift $\Delta^{(\mu )}_{FnJI}$ is known [see (\ref{hfs_shift})], then the coupling matrix element in (\ref{matr_el}) between states with the same electronic configuration $|F,m_{F},(n,J),I\rangle$ and $|F,m_{F},(n,J),I'\rangle$, differing only in the nuclear spin ($I'$\,$\neq$\,$I$), can be expressed as follows
\begin{eqnarray}\label{nediag_el_mu}
&&(-1)^{F+I+J}
\left\{\begin{array}{ccc}
I' & I & 1 \\
J& J & F \\
\end{array}
\right\}
\mu^{}_{I',I}K^{}_{nJ,nJ}= \\
&& \Delta^{(\mu )}_{FnJI}
\frac{\left\{\begin{array}{ccc}
I' & I & 1 \\
J& J & F \\
\end{array}
\right\}}
{\left\{\begin{array}{ccc}
I & I & 1 \\
J& J & F \\
\end{array}
\right\}}
\frac{\mu^{}_{I',I}}{\mu^{}_{I,I}}\,.
\nonumber
\end{eqnarray}

All the above can be also referred to the matrix elements from the quadrupole contribution $\hat{V}^{(\rm hf)}_{\rm Q}$ in the hyperfine interaction operator, for which, in accordance with the known formulas of the quantum theory of angular momentum (see \cite{Varshalovich_1988}), the following expression is true
\begin{eqnarray}\label{Q}
&& \langle F',m_{F'},(n',J'),I'| \hat{V}^{(\rm hf)}_{\rm Q}|F,m^{}_{F},(n,J),I\rangle = \\
&& \delta^{}_{F'F}\delta^{}_{m_{F'}m_{F}}(-1)^{F+I+J'}
\left\{\begin{array}{ccc}
I' & I & 2 \\
J& J' & F \\
\end{array}
\right\}
Q^{(\rm n)}_{I',I}Q^{(\rm e)}_{n'J',nJ}\,, \nonumber
\end{eqnarray}
where $Q^{(\rm n)}_{I',I}$$\equiv$$\langle I'||\hat{\rm Q}^{(\rm n)}||I\rangle$ is the reduced matrix element of the nuclear quadrupole moment $\hat{\rm Q}^{(\rm n)}$ between the nuclear states $|I'\rangle$ and $|I\rangle$ with spin $I'$ and $I$, while $Q^{(\rm e)}_{n'J',nJ}$$\equiv$$\langle n',J'||\hat{\rm Q}^{(\rm e)}||n,J\rangle$ is the reduced matrix element of the quadrupole electron operator $\hat{\rm Q}^{(\rm e)}$ between the states $|n',J'\rangle$ and $|n,J\rangle$. In this case, the electronic states and nuclear states must have in pairs the same parity, and their angular momenta can differ by no more than two: $|J-J'|\leq 2$, $|I-I'|\leq 2$.

In the case of diagonal matrix elements in (\ref{Q}), the value
\begin{equation}\label{hfs_Q}
\Delta^{(\rm Q)}_{FnJI}=(-1)^{F+I+J}
\left\{\begin{array}{ccc}
I & I & 2 \\
J& J & F \\
\end{array}
\right\}
Q^{(\rm n)}_{I,I}Q^{(\rm e)}_{nJ,nJ}\,,
\end{equation}
describes the hyperfine shift of the corresponding hyperfine level due to the quadrupole interaction. In the case of different nuclear states $I$\,$\neq$\,$I'$, the matrix elements in (\ref{Q}) describe the entanglement of states with different nuclear spins, i.e. they also contribute to the electronic bridge together with the magnetic contribution. In particular, if the value of the quadrupole contribution to the hyperfine shift $\Delta^{(\rm Q)}_{FnJI}$ is known [see (\ref{hfs_Q})], then the coupling matrix element in (\ref{Q}) between states with the same electronic configuration $|F,m_{F},(n,J),I\rangle$ and $|F,m_{F},(n,J),I'\rangle$, differing only in the nuclear spin ($I$\,$\neq$\,$I'$), can be expressed as follows
\begin{eqnarray}\label{nediag_el_Q}
&&(-1)^{F+I+J}
\left\{\begin{array}{ccc}
I' & I & 2 \\
J& J & F \\
\end{array}
\right\}
Q^{(\rm n)}_{I',I}Q^{(\rm e)}_{nJ,nJ}= \\
&& \Delta^{(\rm Q)}_{FnJI}
\frac{\left\{\begin{array}{ccc}
I '& I & 2 \\
J& J & F \\
\end{array}
\right\}}
{\left\{\begin{array}{ccc}
I & I & 2 \\
J& J & F \\
\end{array}
\right\}}
\frac{Q^{(\rm n)}_{I',I}}{Q^{(\rm n)}_{I,I}}\,.\nonumber
\end{eqnarray}
Note also, since for the quadrupole contribution the electronic angular momenta of the states $J$ and $J'$ in the general expression (\ref{Q}) can differ by two, this noticeably expands the channels of electron bridge, in relation to the magnetic contribution $\hat{V}^{(\rm hf)}_{\bf B}$, for which the angular momenta of the states $J$ and $J'$ in (\ref{matr_el}) can differ by no more than one.

Let us now consider the specific case of the isotope $^{229}$Th, where we have for the nuclear ground state the spin 5/2 and for the first excited nuclear state the spin 3/2 (isomer $^{229m}$Th). From experimental data (e.g. see \cite{Peik_2021}), we have the values of nuclear $g$-factors $g^{}_{5/2}$=\,0.36 and $g^{}_{3/2}$=$-0.37$, which in accordance with (\ref{diag_el_mu}) leads to the following diagonal reduced matrix elements
\begin{equation}\label{Th_mu}
 \mu^{}_{5/2,5/2}=2.61\mu^{}_{\rm N},\quad  \mu^{}_{3/2,3/2}=-1.43\mu^{}_{\rm N}.
\end{equation}
As for the non-diagonal reduced matrix element $\mu^{}_{3/2,5/2}$, responsible for the entanglement of atomic states with different nuclear spins (i.e. the electron bridge), its value is easily determined by the lifetime $T$ of the nuclear isomer $^{229m}$Th with the spin 3/2, which is about 2000~sec (see \cite{Katori_2024}). For this purpose, we use the known formula for the spontaneous decay rate of the upper level with the spin $I'$ to the lower level with the spin $I$ in the magneto-dipole approximation:
\begin{equation}\label{gamma}
\frac{1}{T}=\frac{32\pi^3|\mu^{}_{I',I}|^2}{3\hbar\lambda^3 (2I'+1)}\,.
\end{equation}
Assuming in this formula $T$\,=\,2000\,sec, $I'$=3/2, $I$=5/2 and $\lambda$=148.38\,nm, we obtain the value
 \begin{equation}\label{mu_m}
\mu^{}_{3/2,5/2}=0.9\mu^{}_{\rm N}.
 \end{equation}
 Note that in contrast to the matrix element $\mu^{}_{3/2,5/2}$, the non-diagonal reduced matrix element $Q^{(\rm n)}_{3/2,5/2}$ for the nuclear quadrupole moment is very difficult to measure directly. However, from general physical considerations, its magnitude should be comparable to the magnitude of the diagonal elements $Q^{(\rm n)}_{5/2,5/2}$ ($\sim$\,3.15\,eb) and $Q^{(\rm n)}_{3/2,3/2}$ ($\sim$\,1.74\,eb), experimentally determined from measurements of the hyperfine level splitting (see \cite{Peik_2021}). Therefore, the contribution to the electron bridge from the nuclear quadrupole moment cab be comparable with the contribution from the nuclear magnetic moment.

As is well known, an electron bridge can lead to a significant acceleration of the spontaneous decay of the isomeric state $^{229m}$Th (see \cite{Tkalya_1991}), which is due to the partial removal of the ban on the dipole ($E1$) transitions between atomic states with different nuclear spins. However, it also opens new possibilities to create the optical clock based on the 148.38~nm nuclear transition in the $^{229}$Th isotope using two-photon spectroscopy, when the clock laser has a doubled wavelength of 296.76~nm.

\begin{figure}[t]
\centerline{\scalebox{0.73}{\includegraphics{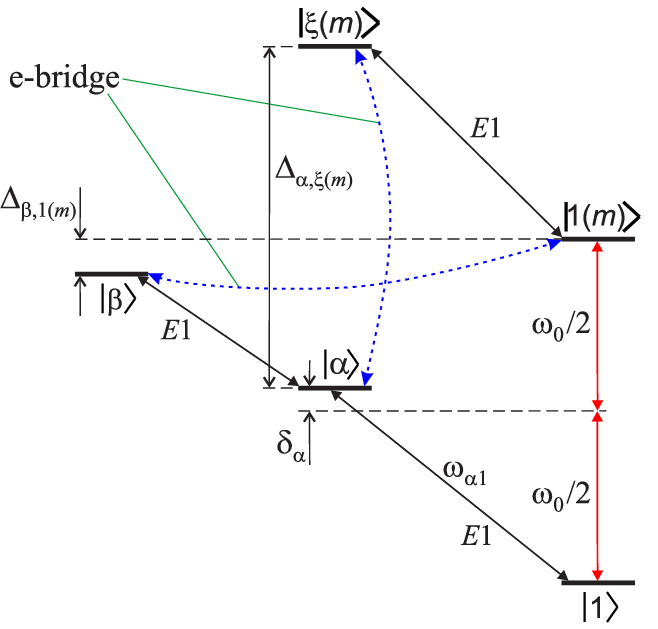}}}\caption{Schematic illustrating one of the possible channels for two-photon spectroscopy of the clock nuclear transition $|1\rangle \leftrightarrow |1(m)\rangle$.
The blue dashed arrows indicate the bonds caused by the electron bridge (e-bridge).} \label{fig1}
\end{figure}

Let us consider possible channels for such spectroscopy using the scheme presented in Fig.\,\ref{fig1}. Here the states $|1\rangle$, $|\alpha\rangle$ and $|\beta\rangle$ are some selected energy states of the ion with the ground nuclear state ($I$=5/2 in the case of the isotope $^{229}$Th), while the upper states $|1(m)\rangle$ and $|\xi(m)\rangle$ are states with the excited nuclear state ($I$=3/2 for $^{229m}$Th). It is assumed that the transition $|1\rangle \leftrightarrow |1(m)\rangle$ is a clock transition with frequency $\omega_0$ ($\omega_0$=$2\pi\times 2020.41$~THz for $^{229}$Th). It is also assumed that the transitions $|1\rangle \leftrightarrow |\alpha\rangle$, $|\alpha\rangle \leftrightarrow |\beta\rangle$ and $|1(m)\rangle \leftrightarrow |\xi(m)\rangle$ are optical dipole ($E1$) transitions. Thus, the parity of the states $|1\rangle$, $|1(m)\rangle$ and $|\beta\rangle$ is opposite to the parity of the states $|\alpha\rangle$ and $|\xi(m)\rangle$.

Under the influence of mixing between states (see blue dashed arrows in  Fig.\,\ref{fig1}) due to the electron bridge, in our case caused by the hyperfine interaction operator $\hat{V}^{(\rm hf)}$, the following transformation of the ion (atom) states $|1(m)\rangle$ and $|\alpha\rangle$, which are of interest to us
\begin{eqnarray}\label{new_states}
|1(m)\rangle &\Rightarrow &|1(m)'\rangle=|1(m)\rangle  +u_\beta|\beta\rangle ,\\
|\alpha\rangle &\Rightarrow &|\alpha'\rangle=|\alpha\rangle +w_\xi |\xi(m)\rangle,\nonumber
\end{eqnarray}
where the small mixing parameters $|u_\beta, w_\xi|$$\ll$1, according to the perturbation theory, are defined as follows
\begin{equation}\label{param}
u_\beta=-\frac{\langle\beta |\hat{V}^{(\rm hf)}|1(m)\rangle}{\hbar\Delta_{\beta,1(m)}},\quad w_\xi=\frac{\langle \xi (m)|\hat{V}^{(\rm hf)}|\alpha\rangle}{\hbar\Delta_{\alpha,\xi(m)}},
\end{equation}
where $\hbar\Delta_{\beta,1(m)}$ is the energy difference between the states $|\beta\rangle$ and $|1(m)\rangle$, while $\hbar\Delta_{\alpha,\xi(m)}$ is the energy difference between the states $|\alpha\rangle$ and $|\xi(m)\rangle$ (see Fig.\,\ref{fig1}).

We consider the interaction of an ion (atom) with a monochromatic laser field
\begin{equation}\label{E}
E(t)={\cal E}_0 e^{-i\omega t}+c.c.\,,
\end{equation}
where ${\cal E}_0$ is the field amplitude and its frequency $\omega$, which is varied near the half frequency of the clock transition $|1\rangle$\,$\leftrightarrow$\,$|1(m)\rangle$, i.e., $\omega$\,$\approx$\,$\omega_0/2$. We will consider the interaction of light with the ion as electro-dipole $-(\hat{d}E)$, where $\hat{d}$  is the dipole moment operator.

Taking into account the mixing of states in (\ref{new_states}), let us consider the two-photon excitation of the clock transition $|1\rangle$\,$\leftrightarrow$\,$|1(m)'\rangle$ at twice the frequency of the laser field $2\omega$\,$\approx$\,$\omega_0$, realized via the intermediate level $|\alpha'\rangle$. In this case, the two-photon Rabi frequency of this excitation is defined in the standard way:
\begin{eqnarray}\label{2ph_rabi}
\frac{\langle 1(m)'|\hat{d}{\cal E}_0|\alpha'\rangle\langle \alpha'|\hat{d}{\cal E}_0|1\rangle}{\hbar^2\delta_\alpha}&=&\frac{{\cal E}^2_0d^{}_{\alpha 1}}{\hbar^2\delta_\alpha}(d^{}_{\beta\alpha}u^*_\beta +d^{(m)*}_{\xi 1}w^{}_\xi)= \nonumber\\
&=&\Omega_{\alpha,\beta}^\text{(1)}+\Omega_{\alpha,\xi (m)}^\text{(2)},
\end{eqnarray}
where $\delta_\alpha$\,=\,($\omega_0/2$\,$-$\,$\omega_{\alpha 1}$) is the one-photon detuning (see Fig.\,\ref{fig1}), $d_{\alpha 1}$\,=\,$\langle \alpha|\hat{d}|1\rangle$, $d_{\beta\alpha}$\,=\,$\langle \beta|\hat{d}|\alpha\rangle$ and $d^{(m)}_{\xi 1}$\,=\,$\langle \xi (m)|\hat{d}|1(m)\rangle$ are the matrix elements of the dipole moment operator for the transitions $|1\rangle$\,$\leftrightarrow$\,$|\alpha\rangle$, $|\alpha\rangle$\,$\leftrightarrow$\,$|\beta\rangle$ and $|1(m)\rangle$\,$\leftrightarrow$\,$|\xi (m)\rangle$, respectively. In (\ref{2ph_rabi}), we took into account only the linear contributions in small parameters ($u_\beta,w_\xi$) and split the expression into two contributions
\begin{equation}\label{2_contr}
\Omega_{\alpha,\beta}^\text{(1)}=\frac{{\cal E}^2_0d^{}_{\alpha 1}}{\hbar^2\delta_\alpha}d^{}_{\beta\alpha}u^*_\beta ,\quad \Omega_{\alpha,\xi (m)}^\text{(2)}=\frac{{\cal E}^2_0d^{}_{\alpha 1}}{\hbar^2\delta_\alpha}d^{(m)*}_{\xi 1}w^{}_\xi .
\end{equation}
Since there are several choices of levels $|\beta\rangle$ and $|\xi(m)\rangle$, the total two-photon Rabi frequency realized via the intermediate level $|\alpha\rangle$, is defined as a superposition over all possible variants
\begin{equation}\label{2ph_alpha}
\Omega^\text{(2-ph)}_{\alpha}=\sum_{\beta}\Omega_{\alpha,\beta}^\text{(1)}+\sum_{\xi}\Omega_{\alpha,\xi (m)}^\text{(2)}.
\end{equation}
Then the total two-photon Rabi frequency for the clock transition $|1\rangle \leftrightarrow |1(m)\rangle$ is the sum over all possible intermediate levels $|\alpha\rangle$
\begin{equation}\label{2ph_total}
\Omega^\text{(2-ph)}_{\rm clock}=\sum_{\alpha}\Omega^\text{(2-ph)}_{\alpha},
\end{equation}
where two-photon transitions via intermediate states $|\alpha (m)\rangle$ with the excited nuclear state are also taken into account.

\begin{figure}[t]
\centerline{\scalebox{0.43}{\includegraphics{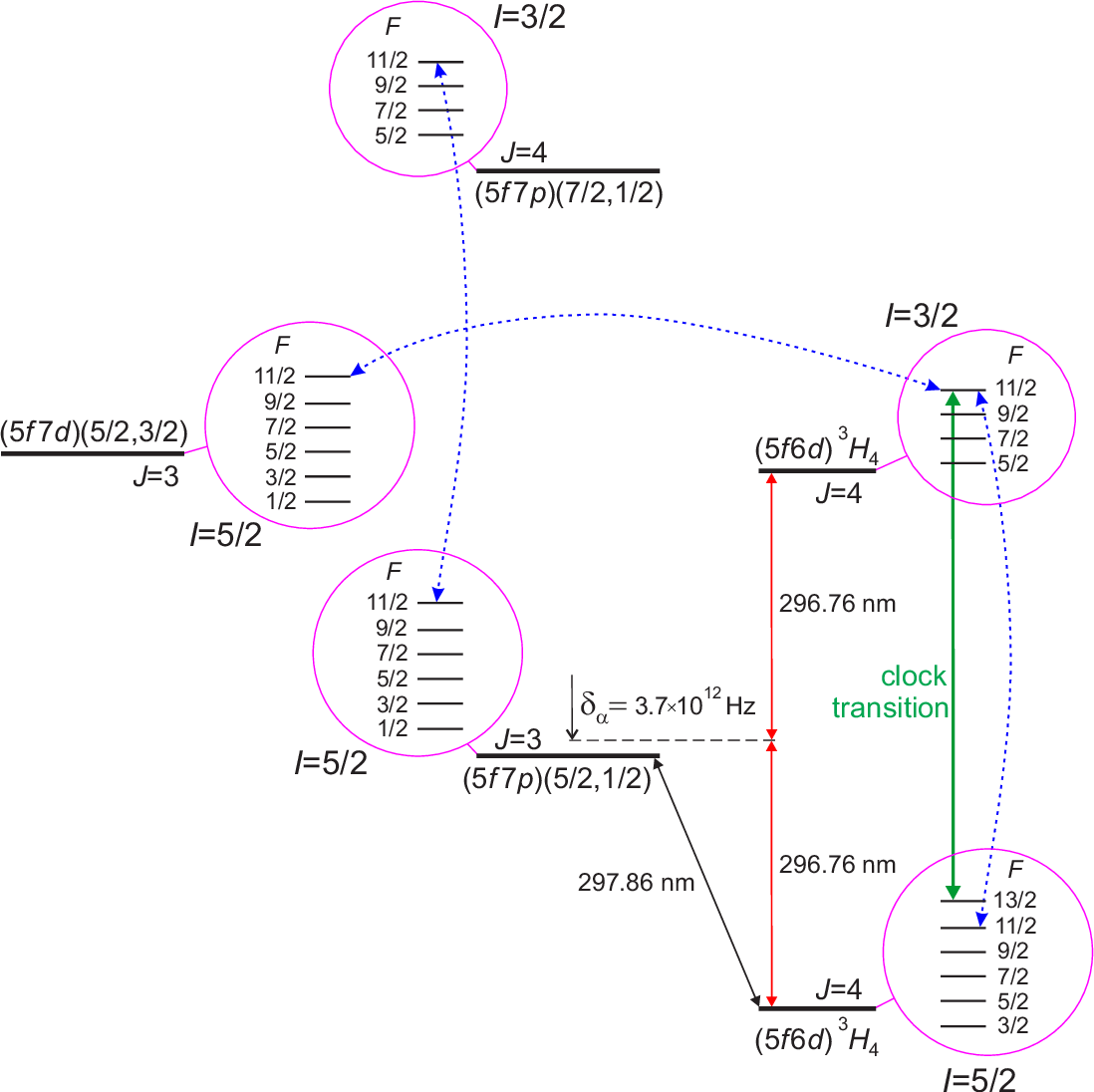}}}\caption{Schematic illustration of the strongest channel for two-photon spectroscopy of the clock nuclear transition in the $^{229}$Th$^{2+}$ ion. For convenience, the arrangement of the hyperfine energy levels (see in the circular tabs) is ordered according to the magnitude of the total angular momentum $F$. The blue dashed arrows mark the electron-bridge bonds for the states, which cause the two-photon spectroscopy of the selected clock transition (see the solid green arrow).} \label{fig2}
\end{figure}

Let us now analyze the possibility of two-photon spectroscopy of the nuclear clock transition for the ion $^{229}$Th$^{2+}$. For this ion, the electronic structure of the ground state is defined as ($5f6d$)\,$^3$$H$$_4$ with an electronic angular momentum $J$=4, which is formed by $LS$-coupling of two valence electrons on the outer electron shells. Following the ideology of \cite{Campbell_2012}, it is reasonable to choose the transition 13/2\,$\leftrightarrow$\,11/2 between hyperfine levels with maximal total angular momenta as the clock nuclear transition (see the transition marked by the green solid arrow in Fig.\,\ref{fig2}). In this case, analyzing the known structures of electronic levels \cite{NIST} for $^{229}$Th$^{2+}$, we have found the following channels (see Fig.\,\ref{fig2}), which contain dominant contributions to the value of the two-photon Rabi frequency (\ref{2ph_total}). These channels are based on an intermediate level (an analog of $|\alpha\rangle$ state in Fig.\,\ref{fig1}) with electronic configuration $(5f7p)\,(5/2,1/2)$ and electronic angular momentum $J$=3, formed by the $jj$-coupling for two valence electrons. The exclusivity of these channels for two-photon spectroscopy of the clock nuclear transition is due to two reasons. First, the exceptional nearness of the wavelength (297.86~nm) of the transition [($5f6d$)\,$^3$$H$$_4$]$\to$[$(5f7p)\,(5/2,1/2)\,J$=3] to the wavelength of the probe laser (296.76~nm), which determines very small one-photon detuning $\delta_\alpha$\,=\,$3.7$$\times$$10^{12}$~Hz and, therefore, a large increase of the corresponding Rabi frequency (\ref{2ph_rabi}). In addition, as follows from \cite{NIST}, this dipole transition [($5f6d$)\,$^3$$H$$_4$]$\to$[$(5f7p)$$(5/2,1/2)\,J$=3] is one of the strongest optical transitions in the ion $^{229}$Th$^{2+}$, which causes a large value of the reduced matrix element of the dipole moment [see $d_{\alpha 1}$ in (\ref{2ph_rabi})].

In this case, as an analog of $|\beta\rangle$ state (see Fig.\,\ref{fig1}), the lower hyperfine level [$I$=5/2,($5f6d$)\,$^3$$H$$_4$,$F$=11/2] is already automatically present, coupled by an electronic bridge (see the right blue dashed arrow in Fig.\,\ref{fig2}) with the upper clock isomeric state  [$I$=3/2,($5f6d$)\,$^3$$H$$_4$,$F$=11/2], which is necessary to form the contribution $\Omega_{\alpha,\beta}^\text{(1)}$ to the two-photon Rabi frequency of the clock transition [see formulas (\ref{2ph_rabi}) and (\ref{2_contr})]. In addition, as another level, which is similar to the state $|\beta\rangle$ in Fig.\,\ref{fig1}, we presented the state [$I$=5/2,$(5f7d)$$(5/2,3/2)$\,$J$=3] with the electronic configuration $(5f7d)\,(5/2,3/2)$ and electron angular momentum $J$=3. This level is distinguished by the relative nearness to the upper clock isomeric state [$I$=3/2,\,($5f6d$)\,$^3$$H$$_4$], which, in turn, can lead to some increase of the value $u_\beta$ in the formula (\ref{2ph_rabi}) for this contribution. Moreover, such states with the nuclear spin $I$=5/2, which are relatively close to the upper clock level [$I$=3/2,\,($5f6d$)\,$^3$$H$$_4$], are quite numerous. This, due to the accumulation effect, can also lead to an increase of the total two-photon Rabi frequency (\ref{2ph_total}). However, one should also take into account the possibility of destructive interference from some contributions, for which the sign of the corresponding two-photon Rabi frequency $\Omega_{\alpha,\beta}^\text{(1)}$ is opposite to the sign of the total two-photon frequency $\Omega^\text{(2-ph)}_{\rm clock}$.

As an analog of the intermediate isomeric state $|\xi(m)\rangle$ (see Fig.\,\ref{fig1}) with the spin of the nucleus $I$=3/2, we have presented in Fig.\,\ref{fig2} the state with the electronic configuration $(5f7p)\,(7/2,1/2)$ and the electronic angular momentum $J$=4, where there is a necessary hyperfine level with the total angular momentum $F$=11/2, coupled by the electron bridge (see blue dashed arrow in Fig.\,\ref{fig2}) with the state [$I$=5/2,$(5f7p)$$(7/2,1/2)$$J$=3,$F$=11/2], which is necessary for the formation of the $\Omega_{\alpha,\xi (m)}^\text{(2)}$ contribution to the two-photon Rabi frequency [see formulas (\ref{2ph_rabi}) and (\ref{2_contr})]. In the ion $^{229}$Th$^{2+}$, there are also several such states, similar to the intermediate isomeric state $|\xi(m)\rangle$ in Fig.\,\ref{fig1}.

Based on the above approach to the description of the electron bridge through the hyperfine interaction operator (\ref{hfs_gen}), we can estimate the magnitude of the coupling matrix elements between ion (atom) states with different values of the nuclear spin $I$ at the level of the magnitude of the hyperfine level splitting. For the ion $^{229}$Th$^{2+}$, this corresponds to the gigahertz order of magnitude (in frequency units). As a result, analyzing the structure of energy levels (see \cite{NIST}) and using formulas (\ref{param}),\,(\ref{2ph_rabi})-(\ref{2ph_total}), we estimate that the intensity at the level of $\sim$\,10-100~kW/cm$^2$ for laser field with wavelength of 296.76~nm can be sufficient to achieve the value of the two-photon Rabi frequency $\Omega^\text{(2-ph)}_{\rm clock}$\,$\sim$\,10\,Hz for the nuclear clock transition with wavelength of 148.38~nm in the ion $^{229}$Th$^{2+}$. Note that 1~W of the light beam power, focused to a diameter of 100~$\mu$m, gives an intensity of about 10~kW/cm$^2$. At the same time, lasers with such a wavelength ($\sim$\,300~nm) and power $\gtrsim$\,1~W are quite accessible.

It can also be noted that the presence of an exceptionally strong (described above) resonance channel for two-photon spectroscopy of the nuclear transition in the ion $^{229}$Th$^{2+}$ leads, on the other hand, to an increase in the residual light shift of the clock transition frequency, which, according to our estimates, can exceed the value of the two-photon Rabi frequency $\Omega^\text{(2-ph)}_{\rm clock}$ by two orders of magnitude. For example, at $\Omega^\text{(2-ph)}_{\rm clock}$\,$\sim$\,10~Hz the light shift can reach the level (and even more) of 1~kHz. However, this problem has a comprehensive solution using the hyper-Ramsey spectroscopy \cite{Yudin_2010,Huntemann_2012} and its modifications \cite{Hobson_2016,Zanon_2018}, as well as the method of autobalanced Ramsey spectroscopy \cite{Sanner_2018,Yudin_2018}. In this case, the laser field ${\bf E}(t)$=Re\{${\bf E}_0e^{-i\omega t}$\} should have linear polarization, the vector of which ${\bf E}_0$ is directed at an angle of 45$^{\circ}$ to the vector of the static magnetic field ${\bf B}_{\rm stat}$, required for the splitting of Zeeman sublevels at the clock hyperfine levels $|F$=13/2$\rangle$ and $|F'$=11/2$\rangle$ (see Fig.\,\ref{fig2}). In this case, the used reference transitions between extreme Zeeman sublevels $|F$=13/2,\,$m$=$\pm$13/2$\rangle$\,$\leftrightarrow$\,$|F'$=11/2,\,$m'$=$\pm$11/2$\rangle$ have the same light shift (due to the linear polarization ${\bf E}_0$), and the corresponding two-photon Rabi frequency $\Omega^\text{(2-ph)}_{\rm clock}$\,$\propto$\,$\sin2\theta$ becomes maximal (here $\theta$ is the angle between the vectors ${\bf E}_0$ and ${\bf B}_{\rm stat}$).

Thus, the possibility of creating a precision optical clock using two-photon spectroscopy of the nuclear transition in the ion $^{229}$Th$^{2+}$ looks quite promising. However, this also requires that the lifetime of the upper isomeric state of the clock [$I$=3/2,($5f6d$)\,$^3$$H$$_4$] be at least 10~seconds to ensure the necessary time duration of the clock transition interrogation. Analysing the energy structure of electronic levels in the ion $^{229}$Th$^{2+}$ and estimating (in accordance with the above theory) the channels for the electron bridge (which lead to an increase in the spontaneous decay rate of the state [$I$=3/2,($5f6d$)\,$^3$$H$$_4$]), we believe that the required lifetime (i.e. not less than 10~sec) is quite possible. However, the final conclusion can be made only after direct experimental measurements. It should be noted that in the case where the decay of the upper isomeric state is mainly determined by the electron bridge, the lifetime of the levels in the hyperfine multiplet can significantly depend on the value of the total angular momentum $F$. In this context, we are primarily interested in the lifetime of the clock upper hyperfine state with $F=11/2$. In addition, for the $^{229}$Th$^{2+}$ ion, other higher and  long-lived electronic levels can be considered as clock levels, for example: ($5f6d$)\,$^3$$H$$_{5,6}$ with $J$=5,\,6 (instead of the lower electronic state ($5f6d$)\,$^3$$H$$_4$), and the possibility of two-photon spectroscopy of the nuclear transition in this case can also be investigated.

As for the ion $^{229}$Th$^{+}$, there is a large density of electronic energy states and, therefore, two-photon spectroscopy of the nuclear transition at an intensity of $\sim$\,10-100~kW/cm$^2$ for laser field with a wavelength of 296.76~nm is also quite probable. However, due to the large density of states, the lifetime of the upper clock isomeric state can be noticeably less than 10~sec, because of the amplification and multiplicity of different decay channels due to the electron bridge. Although, the final conclusion can be made only after direct experimental measurements. As for the ion $^{229}$Th$^{3+}$, on the contrary, there is very large lifetime of the upper clock isomeric state, which is about 2000~sec \cite{Katori_2024}. However, analysing the electronic levels presented in \cite{Safronova_2013}, we did not find a dedicated resonance channel for two-photon spectroscopy (similar to the channel for the $^{229}$Th$^{2+}$ shown in Fig.\,\ref{fig2}). Therefore, for the ion $^{229}$Th$^{3+}$, we estimate that the required intensity will much exceed 100~kW/cm$^2$ for the laser field with a wavelength of 296.76~nm, which can be a significant technical problem.

In conclusion, for the isotope $^{229}$Th we investigated the possibility of two-photon laser spectroscopy of the nuclear clock transition (148.38~nm) using intense monochromatic laser field at twice wavelength (296.76~nm). Our estimates show that using the electron bridge mechanism in the doubly ionized ion $^{229}$Th$^{2+}$ the sufficient intensity of a continuous laser field will be of the order of 10-100~kW/cm$^2$, which lies within the reach of existing modern laser systems. For example, a laser beam with a power of 1~W, focused into a diameter of 100~$\mu$m, gives an intensity of about 10~kW/cm$^2$. In the case of experimental confirmation of our predictions, the results obtained can be the basis for the practical creation of ultra-precision nuclear optical clock based on thorium-229 ions, trapped in the ion trap, without using vacuum ultraviolet. This may also stimulate the search for solid-state crystals with a transparency window in the region of 296.76~nm, in which the doped thorium ions would be in the charge state $^{229}$Th$^{2+}$. In this case, it would be possible to create solid-state clocks based on two-photon laser spectroscopy of the nuclear clock transition. Obviously, in a such solid-state device the laser field intensity should not be so high (10-100~kW/cm$^2$), since the necessary level of spectroscopic signal will be provided by a large number of thorium ions in the crystalline sample.

Moreover, we develop an alternative approach to the description of the electron bridge phenomenon in an isolated ion (atom) using the hyperfine interaction operator, that is important for the general quantum theory of an atom. In particular, this approach shows that the contribution to the electron bridge from the nuclear quadrupole moment can be comparable, in the general case, to the contribution from the nuclear magnetic moment. Such an approach can also be extended to the case of ions doped in a solid.

We thank M.~D.~Radchenko for useful  discussions.

\end{document}